\documentclass[conference]{IEEEtran}
\usepackage[T1]{fontenc}% optional T1 font encoding

\ifCLASSINFOpdf
\else
\fi
\usepackage{amsmath,amssymb} % amssymb is for multiline equations in a table
\usepackage[cmintegrals]{newtxmath}
\usepackage{url} %To make URL appears in bibliography
\usepackage{blindtext, graphicx}
\usepackage[font=small,skip=0pt]{caption}
\usepackage{float} %to make tables and figures appear at specified location
\usepackage{multirow}
\usepackage{array}
 
\usepackage{multicol}

\usepackage{cite}

\usepackage{amsthm}

\usepackage{flushend} % needed to make last page columns balanced rather than empty right column and filled left one

\IEEEoverridecommandlockouts %needed to make thanks appear (I need to write it at the bottom of the first page)
\begin{document}
\title{Closed Form Expressions for the Probability Density Function of the Interference Power in PPP Networks}

%\author{Author~1, Author~2, Author~3, Author~4, Author~5
%}
\author{Hussein~A.~Ammar, Youssef~Nasser, Hassan~Artail\thanks{This work will be presented at the IEEE International Conference on Communications (ICC'18).}\\
	Department of Electrical and Computer Engineering\\American University of Beirut, Beirut 1107 2020, Lebanon\\
	\{haa141, yn10, hartail\}@aub.edu.lb
}

\maketitle


%\footnote{This work will be presented at the IEEE International Conference on Communications (ICC'18)}
\begin{abstract}
In this paper, we provide closed form expressions for the probability density functions (PDF) of the interference power in a network whose transmitters are arranged according to the Poisson Point Process (PPP). These expressions apply for any integer path loss exponent $\eta$ greater than 2. Using the stretched exponential or Kohlrausch function, we show that the PDF formulas can be obtained as long as the Laplace transform (LT) for the PDF follows a specific common (exponential) formulation. Moreover, we show that such closed form expressions can be useful in deriving performance metrics for the network for any fading type experienced by the signals. Finally, using Monte-Carlo simulations and numerical analysis, we validate the accuracy of the proposed analytical derivations.
\end{abstract}


% \fi
\IEEEpeerreviewmaketitle

\section{Introduction}
%\IEEEPARstart{T}{his} demo file is
%In wireless networks, operators spend huge efforts on their network planning in order to provide better services and performance to their customers. Nonetheless, these efforts are more effective when the operators have reliable analytical formulation for the network performance metrics. The derivation of the interference analytical models, distribution, and characteristics, will therefore give network planners the needed tools for deriving the network performance in different scenarios and conditions. Additionally, as the wireless networks are becoming increasingly denser and experiencing higher traffic load, as defined by the connection density and area traffic capacity in the requirements of 5G access technologies \cite{my5GTechnicalManual}, the need to characterize interference becomes more critical.
\par In wireless networks, operators spend huge amount of effort on their network planning to provide better services and performance to their customers. The network planning process consists of many phases related to modeling, managing and tuning the network or the cells configuration, in order to increase the efficiency of the network and achieve the needed quality of service. Nonetheless, these efforts are more efficient when the operators have a solid knowledge and reliable analytical information relating to the network performance metrics. In practice, the signal to interference and noise ratio (SINR) received by users, plays a major role in determining the network performance. Particularly, the interference is a main component affecting the SINR, and hence the operators need to better understand its characteristics.
\par In stochastic geometry, the Laplace Transform (LT) for the interference power distribution is used extensively to get many performance metrics for the network \cite{5226957}. Stochastic geometry has been widely applied to derive the probability of coverage and capacity expressions, among many others. It is a common tool for network evaluation, although it requires extensive mathematical derivations. The derivation of many performance metrics has been based on the analytical expressions of the LT of the interference power. While the LT gives a good understanding of the interference, it might hide the shape and form of its probability distribution function (PDF), which in turn gives an extensible method for the network operators to understand the interference behavior and trends (distribution shape, lower-bound, upper bound, and other possible statistics). The Poisson Point Process (PPP) network is the most common model in network analysis using stochastic geometry. In the literature, PPP has been shown that it provides a good representation of real network deployments (whether it fits to derive the real performance, or a pessimistic one). It is widely known that when transmitters locations are arranged as PPP, there is no closed form expressions for the PDF of the interference power, except for the specific case of path loss exponent (PLE) $\eta=4$ \cite{haenggi2009interference}. As a result, few approximations have been provided in this regard, as described in the next section, but they are limited to particular transmission scenarios.

\subsection{Literature Review}
\par In the literature, and particularly in networks whose transmitters' locations are arranged according to PPP, the approaches used to characterize the interference power, range from treating the LT of interference, to providing the characteristic and the moment generating functions. Some techniques may even derive further network performance metrics, like the probability of coverage provided by the network, without fully characterizing the interference from all sources in the network. Such techniques calculate the interference from the $n^\text{th}$ nearest sources only. But mainly, a lot of studies stick with the assumption that the interference and the useful signal is subject to Rayleigh fading only, thus allowing the use of the LT of the interference to characterize further system performance metrics.
\par The authors of \cite{6524460} gave a summary of the main approaches used to characterize the aggregate interference, where different approximations with simple PDFs are used to approximate the interference power. In \cite{7733098}, the authors analyzed the use of the characteristic function of interference to derive its moments. Also, with the aid of the numerical analysis, they studied the interference distribution in the presence and absence of an interference exclusion region around the studied user, which seems to change the interference power distribution. Others have used different approximations for the PDF of interference power \cite{6515339,6189009,5962648,4133877}, which was fitted with distributions having closed form PDF expressions. This was mainly done through moment matching methods, where the error between the true empirical PDF and the approximated one was analyzed using simulations. Moreover, numerical inversion of the LT or the characteristic function was used \cite{5935092}. To the best of the authors' knowledge, the general PDF of the interference power in PPP networks has not been addressed, that is, except for the case mentioned above and given in \cite{haenggi2009interference}, in which the path loss exponent (PLE) is restricted to 4. And it is widely known in the literature that, there is no known expression for the pdf of the aggregate interference in PPP networks.
%Others tried to characterize interference using different methods other than starting from the Laplace domain [], in which .
\subsection{Importance and Contribution}
\par In this paper, we state that exact closed form PDF expressions for the interference power in any PPP network model configuration exist as long as we can write their formulas in the Laplace domain in a specific exponential structure. Our main contribution lies in showing that the stretched exponential function is a suitable choice for obtaining the formulas of the interference distribution. We illustrate how this distribution can be written in terms of a modified L\'evy distribution, a specific type of alpha-stable functions.% We also show how the Generalized Extreme Value (GEV) distribution can be used as an approximation for the interference power.
\par The work we present is important for performing further network performance analysis, and specifically for directly obtaining the interference statistics and thus deriving the SINR in PPP networks. This in turn helps in understanding network enhancement techniques that can better tune the interference distribution. Besides, a compact closed form expression allows for plugging parameters and deriving network performance measures in an infinite number of scenarios. For instance, different distributions for the channel fading on the useful signal and interference signals could be considered. 

To our knowledge, this is the first work dealing with exact expressions for the interference PDFs for any integer value of the PLE.
\par The rest of the paper is organized as follows. Section II introduces a simple mathematical background about the stretched exponential function. Section III analyzes the interference power in PPP networks under a stochastic geometry framework, and investigates the applicability and the accuracy of the presented stretched exponential functions. Section IV shows one use of the interference power PDF in deriving the probability of coverage for any fading experienced by the signals. Finally, we present our conclusion in Section V.
%The rest of the letter is organized as follows. Sections II and III introduce a simple mathematical background about the stretched exponential function, and investigate its applicability in interference analysis in a stochastic geometry framework, and accordingly we show the accuracy of the presented expressions. In section IV, we show the feasibility of using GEV as an approximation, in which we can write simpler forms for the interference for the interference power. Finally, we present our conclusion in Section V.
\section{Mathematical Preliminaries}
The stretched exponential function, %which is the Kohlrausch function,
or the Kohlrausch-Williams-Watts (KWW) function \cite{anderssen2004kohlrausch}, %\cite{kohlrausch1854theorie}
is defined as:
\begin{align}\label{eq:KWW}
F_\beta (s)=e^{-s^\beta}
\end{align}
%To get the unilateral inverse Laplace transform (LT) of this function, we use the definition established by Pollard \cite{pollard1946representation} that uses the index $\beta$, and is written in the form:
It is directly related to the Laplace domain of the Levy distribution as:
\begin{align}
L_{f_\beta}(s)=\int_{0}^{\infty}e^{-sI}f_\beta(I)d I=e^{-s^\beta}
\end{align}
where $f_\beta(I)$ is a stable PDF having a stretching exponent $\beta$ such that $0<\beta<1$. The $\beta$ exponent is usually considered as a ratio term such $\beta=\frac{\beta_1}{\beta_2}$ where $\beta_1$ and $\beta_2$ are integers. For $\beta_1=1$ and $\beta_2=2$, the inverse LT (ILT) of the KWW function leads to the simplest PDF expression, known as L\'evy distribution\footnote{Some authors use the term L\'evy distribution for all sum stable laws}, given by:
\begin{align}\label{eq:beta2}
f_{\frac{1}{2}}(I)=\frac{exp\left(-\frac{1}{4 I}\right)}{2\sqrt{\pi}I^{\frac{3}{2}}}
\end{align}
\begin{table*}[ht]
	\centering
	\begin{tabular}[t]{|p{0.01\linewidth}|p{0.4\linewidth}|}
		\hline
		$\beta$ &PDF of $I$=\\
		\hline
		\hline
		\vspace{1mm}
		$\frac{1}{2}$
		&
		\vspace{-2.3em}
		\begingroup\makeatletter\def\f@size{6}\check@mathfonts
		\begin{align*}%\label{eq:A}
		\frac{1}{t^2}f_\frac{1}{2}\left(\frac{I}{t^2}\right)=\frac{t exp\left(-\frac{t^2}{4 I}\right)}{2\sqrt{\pi}I^\frac{3}{2}}
		\end{align*}
		\endgroup
		\vspace{-1em}
		\\
		\hline
		\multirow{2}{1\linewidth}{
		$\frac{1}{3}$
		} &
		\vspace{-2.3em}
		\begingroup\makeatletter\def\f@size{6}\check@mathfonts
		\begin{align*}%\label{eq:B}
		\dfrac{1}{t^3}f_{\frac{1}{3}}\left(\dfrac{I}{t^3}\right)=\dfrac{t^\frac{3}{2}}{3\pi I^\frac{3}{2}}K_{\frac{1}{3}}\left(\dfrac{2}{3\sqrt{\frac{3 I}{t^3}}}\right)\end{align*}
		\endgroup
		\vspace{-1em}
		\\
		&
		where $K_v(z)$ is the modified Bessel function of the second kind.
		\\ 
		\hline
		\multirow{2}{1\linewidth}{
		$\frac{2}{3}$ } &
		\vspace{-2.5em}
		\begingroup\makeatletter\def\f@size{6}\check@mathfonts
		\begin{multline*}%\label{eq:C}
		%\begin{array} {lcl}
		\frac{1}{t^\frac{3}{2}}f_{\frac{2}{3}}\left(\dfrac{I}{t^\frac{3}{2}}\right)=\frac{2\sqrt{3}t^3}{27\pi I^3}exp\left(-\dfrac{2t^3}{27 I^2}\right)\left(K_{\frac{1}{3}}\left(\dfrac{2t^3}{27 I^2}\right)+K_{\frac{2}{3}}\left(\dfrac{2t^3}{27 I^2}\right)\right)\\
		=\frac{\Gamma\left(\frac{2}{3}\right)t}{\sqrt{3}\pi I^{\frac{5}{3}}} \,_1F_1\left(\frac{5}{6};\frac{2}{3};-\frac{\frac{2^2}{3^3}t^3}{I^2} \right)+
		\frac{\frac{2}{9}t^2}{\Gamma\left(\frac{2}{3}\right)I^\frac{7}{3}}
		\,_1F_1\left(\frac{7}{6};\frac{4}{3};-\frac{\frac{2^2}{3^3}t^3}{I^2} \right)
		%\end{array}
		\end{multline*}
		\endgroup
		\vspace{-1em}
		\\
		&
		where $\,_pF_q(a_1,...,a_p;b_1,...,b_q;c)$ is the generalized hypergeometric function.
		\\
		\hline
		\vspace{1mm}
		$\frac{1}{5}$ &
		\vspace{-2.4em}
		\begingroup\makeatletter\def\f@size{6}\check@mathfonts
		\begin{align*}%\label{eq:E}
		\frac{1}{t^5}f_{\frac{1}{5}}\left(\dfrac{I}{t^5}\right)
		=\frac{1}{t^5}\sum_{m=1}^{4}\frac{b_m\left(5,1\right)}{\left(\frac{I}{t^5}\right)^{1+\frac{m}{5}}} \,_2F_5\left(\left[1,\Delta\left(1,1+\frac{m}{5}\right)\right];\left[\Delta\left(5,1+m\right)\right];\frac{t^5}{5^5 I}\right)
		\end{align*}
		\endgroup
		\vspace{-1em}
		\\
		&
		where $\Delta\left(a,b\right)=\frac{b}{a},\frac{b+1}{a},...,\frac{b+a-1}{a}$ and
		\\
		&
		$b_1(5,1)=\frac{\sqrt{5}\Gamma\left(\frac{1}{5}\right)}{20\pi \text{sin}\left(\frac{2\pi}{5}\right)}$, $b_2(5,1)=\frac{-\sqrt{5}\Gamma\left(\frac{2}{5}\right)}{20\pi \text{sin}\left(\frac{\pi}{5}\right)}$,
		\\
		&
		$b_3(5,1)=\frac{\sqrt{5}\Gamma\left(\frac{3}{5}\right)}{40\pi \text{sin}\left(\frac{\pi}{5}\right)}$, $b_4(5,1)=\frac{-\sqrt{5}\Gamma\left(\frac{4}{5}\right)}{120\pi \text{sin}\left(\frac{2\pi}{5}\right)}$
		\\
		\hline
	\end{tabular}
	\quad\quad\quad
	\begin{tabular}[t]{|p{0.01\linewidth}|p{0.4\linewidth}|} 
		\hline
		$\beta$ &PDF of $I$=\\
		\hline
		\hline
		\vspace{1mm}
		$\frac{1}{4}$ &
		\vspace{-2.5em}
		\begingroup\makeatletter\def\f@size{6}\check@mathfonts
		\begin{multline*}%\label{eq:D}
		\frac{1}{t^4}f_{\frac{1}{4}}\left(\dfrac{I}{t^4}\right)
		= \frac{t^3}{64 \pi I^\frac{7}{4}}\left(\frac{8\sqrt{2 I}}{t^2} \Gamma \left(\frac{1}{4}\right) \, _0F_2\left(;\frac{1}{2},\frac{3}{4};-\frac{t^4}{256 I}\right)-\sqrt{2}\Gamma \left(-\frac{1}{4}\right)\right.\\
		\left.
		\, _0F_2\left(;\frac{5}{4},\frac{3}{2};-\frac{t^4}{256 I}\right)-16 \sqrt{\pi } \dfrac{I^\frac{1}{4}}{t} \, _0F_2\left(;\frac{3}{4},\frac{5}{4};-\frac{t^4}{256 I}\right)\right)
		\end{multline*}
		\endgroup
		\vspace{-1em}
		\\
		\hline
		\vspace{1mm}
		$\frac{2}{5}$ &
		\vspace{-2.5em}
		\begingroup\makeatletter\def\f@size{6}\check@mathfonts
		\begin{multline*}%\label{eq:F}
		\frac{1}{t^\frac{5}{2}}f_{\frac{2}{5}}\left(\dfrac{I}{t^\frac{5}{2}}\right)
		=\frac{1}{t^\frac{5}{2}}\sum_{m=1}^{4}\frac{b_m\left(5,2\right)}{\left(\frac{I}{t^{5/2}}\right)^{1+\frac{2m}{5}}}\\
		\end{multline*}
		%it was divided to make the distance smaller
		\vspace{-5em}
		\begin{multline*} \quad\quad\quad\quad\,_3F_5\left(1,\Delta\left(2,1+\frac{2m}{5}\right);\Delta\left(5,1+m\right);\frac{2^2}{5^5\left(\frac{I}{t^{5/2}}\right)^2}\right)
		\begingroup\makeatletter\def\f@size{8}\check@mathfonts
		\quad\text{where}
		\endgroup
		\end{multline*}
		\endgroup
		\vspace{-1em}
		\\
		&
		$b_1(5,2)=\frac{2^\frac{2}{5}\sqrt{5}\Gamma\left(\frac{1}{5}\right)}{10\sqrt{\pi}\Gamma\left(\frac{3}{10}\right) \text{sin}\left(\frac{2\pi}{5}\right)}$, $b_2(5,2)=\frac{-2^\frac{4}{5}\sqrt{5}\Gamma\left(\frac{2}{5}\right)}{10\sqrt{\pi}\Gamma\left(\frac{1}{10}\right) \text{sin}\left(\frac{\pi}{5}\right)}$,
		\\
		&
		$b_3(5,2)=\frac{-2^\frac{1}{5}\sqrt{5}\Gamma\left(\frac{3}{5}\right)}{100\sqrt{\pi}\Gamma\left(\frac{9}{10}\right) \text{sin}\left(\frac{\pi}{5}\right)}$, $b_4(5,2)=\frac{2^\frac{3}{5}\sqrt{5}\Gamma\left(\frac{4}{5}\right)}{100\sqrt{\pi}\Gamma\left(\frac{7}{10}\right) \text{sin}\left(\frac{2\pi}{5}\right)}$
		\\
		\hline
		\vspace{1mm}
		$\frac{1}{6}$ &
		\vspace{-2.5em}
		\begingroup\makeatletter\def\f@size{6}\check@mathfonts
		\begin{multline*}%\label{eq:G}
		\frac{1}{t^6}f_{\frac{1}{6}}\left(\dfrac{I}{t^6}\right)
		=\frac{2^\frac{-1}{3}3^{\frac{-3}{2}} \sqrt{\pi}t}{\left(\Gamma \left(\frac{2}{3}\right)\right)^2 I^{\frac{7}{6}}}\, _0F_4\left(;\frac{1}{3},\frac{1}{2},\frac{2}{3},\frac{5}{6};-\frac{t^6}{6^6 I}\right)
		-\frac{t^2}{6 \Gamma \left(\frac{2}{3}\right)I^{4/3}}\\
		\,_0F_4\left(;\frac{1}{2},\frac{2}{3},\frac{5}{6},\frac{7}{6};-\frac{t^6}{6^6 I}\right)
		+\frac{t^3}{12\sqrt{\pi }I^{\frac{3}{2}}}\,_0F_4\left(;\frac{2}{3},\frac{5}{6},\frac{7}{6},\frac{4}{3};-\frac{t^6}{6^6 I}\right)
		-\frac{\sqrt{3}t^4 \Gamma \left(\frac{2}{3}\right)}{72 \pi I^{\frac{5}{3}}}\\
		\,_0F_4\left(;\frac{5}{6},\frac{7}{6},\frac{4}{3},\frac{3}{2};-\frac{t^6}{6^6 I}\right)
		+\frac{3^{-\frac{3}{2}}t^5 \left(\Gamma \left(\frac{2}{3}\right)\right)^2}{2^{\frac{17}{3}} \pi ^{\frac{3}{2}} I^{\frac{11}{6}}}\, _0F_4\left(;\frac{7}{6},\frac{4}{3},\frac{3}{2},\frac{5}{3};-\frac{t^6}{6^6 I}\right)
		\end{multline*}
		\endgroup
		\vspace{-1em}
		\\
		\hline
	\end{tabular}
	\vspace{0.5em}
	\caption{Inverse Laplace for scaled KWW $F(s)=exp(-ts^\beta)$ for mostly needed $\beta$ indexes \cite{uchaikin1999chance,penson2010exact,gorska2012levy}.}
	\label{tab:1}
	\vspace{-1em}
\end{table*}
The L\'evy distribution of the random variable $I$ is defined, theoretically, with the following parameters: stability = $0.5$, skewness $=1$, scale $=0.5$, and location $=0$. 
% stability $\alpha$ = scale $\sigma=0.5$, skewness $\epsilon=1$, and location $l=0$.
The interest in the L\'evy distribution resides in its use in the scaled version of the KWW. Indeed, the Inverse unilateral Laplace Transform (ILT) of the scaled version (i.e., $e^{-ts^\beta}$ with scaling parameter $t$) of $F_\beta(s)$ can be obtained through using the time scaling property of the LT.
\par In this paper, we use the main properties of the KWW functions and their ILTs to derive the PDF of the interference power. In \cite{uchaikin1999chance}, different  KWW functions $F_\beta(s)$ have been defined with lower or higher orders of $\beta$ such as $1/3$, $2/3$, $1/4$, $1/5$, $2/5$, etc.\footnote{It should be noted that the only case where the PDF of the interference has been derived is for PLE=4. This is equivalent to the case $\beta=1/2$ in the KWW formulation.} The KWW functions have never been introduced in the stochastic geometry literature for PDF derivations. Here, this is done by using the following important proposition:
\\
\par \textit{\textbf{Proposition}}: The PDF of the interference power can be derived if its LT can be written in a KWW function form.
\begin{proof}
	This is a direct application of the ILT and the scaling properties of the LT, and is discussed in the next section.
\end{proof}
The main problem turns out in finding the ILT of the KWW functions for different values of $\beta$. To do so, we introduce the following property \cite{gorska2012levy}, which allows obtaining the ILT for the $F_\beta(s)$ with higher orders of $\beta$ from lower order ones:
\\
\par \textit{\textbf{Property 1}}: the PDF $f_{\{\beta_a.\beta_b\}}(I)$ can be obtained from $f_{\beta_a}$ and $f_{\beta_a}$ through a simple integration given by:

%\begingroup\makeatletter\def\f@size{6}\check@mathfonts
\begin{align}\label{eq:transitive}
f_{\{\beta_a.\beta_b\}}(I)&=\int_{0}^{\infty}\frac{1}{t^{\frac{1}{\beta_{_a}}}}f_{\beta_{_a}}\left(\frac{I}{t^\frac{1}{\beta_{_a}}}\right)f_{\beta_{_b}}(t)dt\nonumber\\
&=\int_{0}^{\infty}\frac{1}{t^{\frac{1}{\beta_{_b}}}}f_{\beta_{_b}}\left(\frac{I}{t^\frac{1}{\beta_{_b}}}\right)f_{\beta_{_a}}(t)dt
\end{align}
%\endgroup
where $(.)$ means multiplication, and $\beta_{_a}$ and $\beta_{_b}$ are two lower order stable distributions that follow the same rules of $\beta$.
%Using the properties of the LT, we can apply weight scaling to get the Inverse Laplace Transform (ILT) for $e^{-ts^\beta}$ (case for $\beta=1/2$ in TABLE \ref{tab:1}), where $"t"$ can be any shifting constant. 
%In literature and mainly in wireless communications using stochastic geometry approaches, the use of KWW functions and properties can be very wide. However, the analysis has been limited to $\beta=1/2$, which can be solved directly.
%There are many works that proposed explicit forms and procedures for the evaluation of different values of $\beta_1$ and $\beta_2$ \cite{penson2010exact}. One of the most important procedures used to obtain higher orders of $\beta$  from lower orders, is a transitive property for the stable laws defined in \cite{gorska2012levy}:
%where $\beta_{_a}$ and $\beta_{_b}$ are two lower order stable distributions that follow the same rules of $\beta$.
This equation becomes instrumental when getting different values for $\beta$. For example, the formula when $\beta=1/4$ i.e. $f_{\{\beta_a.\beta_b\}}(I)=f_{1/4}(I)$ can be obtained by setting $\beta_a=1/2$ and $\beta_b=1/2$, thus using equation \eqref{eq:beta2}, and substituting it in \eqref{eq:transitive}. As we are interested in the scaled KWW function, we then make time scaling by $\frac{1}{t^\frac{1}{\beta}}$ to obtain the PDF which is the ILT of the KWW i.e. $PDF= \frac{1}{t^{\frac{1}{\beta}}}f_{\beta}\left(\frac{I}{t^{\frac{1}{\beta}}}\right)\leftrightarrow \exp\left(-(t^{\frac{1}{\beta}}s)^\beta\right)=\exp\left(-ts^\beta\right)$ and this leads to the final formula written in Table \ref{tab:1} for the $\beta=1/4$ case. Other higher order formulas have been introduced in \cite{penson2010exact}, where they were written for different values of $\beta$ as a finite sum of generalized Hypergeometric functions $\,_pF_q$.
In Table \ref{tab:1}, we provide the resulting formulations of the KWW literature and the application of \eqref{eq:transitive}, for different important values of $\beta$. In this table, $\beta$ has been selected to match with the common PLE values used in PPP networks.
\vspace{1em}
\section{Derivation and Analysis of the Interference PDF in PPP Network}
%Our numerical analysis of the interference power distribution obtained using PPP stochastic geometry models shows that the interference distribution structure is non-linear and is not purely exponential. This introduces the possibility of using heavy-tailed distributions for modeling the interference power.
We consider a network model in which the transmitters are arranged according to a homogeneous PPP $\mathrm{\Phi}$ with a density $\lambda$ on $\mathbb{R}^2$ with infinite plane. Without loss of generality, we analyze the interference for a receiver taken as the reference/typical user (observation point), located at the origin. According to Slivnyak's theorem \cite{haenggi2012stochastic}, the statistical characteristics seen from a homogeneous PPP are independent of the receiver position. This receiver is experiencing aggregate interference from the transmitters in the network, without the existence of an exclusion area (protection area) for interference around the typical user. 
%This receiver is associated with nearest transmitter $\text{Tx}_\text{serving}$, hence experiencing aggregate interference from other transmitters in the network further than $\text{Tx}_\text{serving}$. 
% For this purpose, we consider two particular cases for interference analysis. The first is the common interference model in PPP \cite{6042301}, while the second is when the base station (BS) distribution has a specific random density $\lambda$.
As a result, the aggregate interference $I$ is defined by:
%\backslash\text{Tx}_{\text{serving}}
\begin{align}\label{eq:interfererence_PPP1}
I=\sum _{i\in\mathrm{\Phi}}g_i{R_i}^{-\eta}
\end{align}
%where $\mathrm{\Phi}$ represents the locations of the BSs that are modeled as a PPP with density $\lambda$,
where $\eta$ is the PLE, $g_i$ is the fading power channel coefficient for arbitrary but identical distributions for all $i$, and $R_i$ is the distance from the typical user to the interfering transmitters which depends on the transmitters' locations that are arranged according to PPP.
\subsection{Laplace Transform of the Interference Power}
\par The analysis of the LTs of the interference power at the receiver in some PPP environments concludes that it can be written as a modified KWW function in which $\beta$ is related to the PLE $\eta$, as will be seen next; hence the importance of the formulas in Table \ref{tab:1}. The LT allows for obtaining many useful metrics for the interference and for the network performance. For example, the probability of coverage $p_c$ for a reference user can be directly obtained from the LT when the useful signal received by this user experiences Rayleigh fading \cite{6042301}. The LT of the interference power received by a typical user in a homogeneous PPP network of transmitters is defined as:
\begin{align}\label{eq:PPP_laplace}
L_I\left(s\right)=\exp{\left(-\pi\lambda\mathbb{E}\left[g^\frac{2}{\eta}\right]\mathrm{\Gamma}\left(1-\frac{2}{\eta}\right)s^\frac{2}{\eta}\right)}=\exp\left(-ts^\beta\right)
\end{align}
\begin{proof}
	See Appendix.% \ref{Appendix}.
\end{proof}
%\begin{align}\label{eq:PPP_laplace}
%L_{I}(s)=exp\left(-\pi\lambda\left(\frac{s}{\mu}\right)^\frac{2}{\eta}\int_{T^{{}^{-2}/{}_\eta}}^{\infty}\frac{1}{y^\frac{\eta}{2}+1}dy\right)
%\end{align}
%where $1/\mu$ is the mean of the fading which is assumed Rayleigh, $T$ is the SINR threshold to guarantee the coverage, and $r$ is the distance to the serving or intended transmitter.\\
where $\Gamma(x)$ is the Gamma function, and $\mathbb{E}[x]$ represents the expectation over the variable $x$.
\par It is clear that Equation \eqref{eq:PPP_laplace} can be written, for any fading distribution, as a KWW function. Therein,  $\beta=2/\eta$, and $t$ is the scaling factor of $s$ which depends on $\lambda,\eta$, and the fading. Hence, its ILT, i.e., the PDF of the interference power, will be a direct plug-in in Table \ref{tab:1}, depending on the value of $\eta$ that determines which equation in the table to use. As for $\mathbb{E}\left[g^\frac{2}{\eta}\right]$, it can be written for different fading distributions as follows.
\\
\\
\textbf{\textit{ LT of the interference in Nakagami fading}}: Nakagami fading is a more general fading distribution whose parameters can be adjusted to a variety of empirical measurements including the Rayleigh and the Rician fading. In this case, the power of fading $g$ is Gamma distributed, i.e., $P_{G}\left(g\right)=\left(\frac{m}{P_r}\right)^m\frac{g^{m-1}}{\mathrm{\Gamma}\left(m\right)}\exp{\left(-\frac{mg}{P_r}\right)}$
, where $m$ is the fading parameter (or the shape of the distribution), and $P_r$ is the average received power ($\frac{m}{P_r}$ is the rate of the distribution). %And they are translated in the Gamma distribution as $\frac{m}{P_r}$ being the rate and $m$ the shape parameter of the distribution. 
Hence, for the Nakagami fading we have:
%\begingroup\makeatletter\def\f@size{8}\check@mathfonts
% needed to don't make the font size of the number of th equation change
%\makeatletter
%\def\tagform@#1{\maketag@@@{\normalsize(#1)\@@italiccorr}}
%\makeatother
\begin{align}
\mathbb{E}^\text{Nakagami}\left[g^\frac{2}{\eta}\right]=\frac{\mathrm{\Gamma}\left(m+\frac{2}{\eta}\right)}{\mathrm{\Gamma}\left(m\right)\left(\frac{m}{P_r}\right)^\frac{2}{\eta}}
\end{align}
%\endgroup

\textbf{\textit{ LT of the interference in Rayleigh fading}}: The Rayleigh fading case is present when there is no Line-Of-Sight (LOS) component in the signal. When $m=1$, the Nakagami fading becomes the Rayleigh fading case. This means that the fading power follows an exponential distribution with mean $\frac{1}{\mu}$. Hence the 
%\begingroup\makeatletter\def\f@size{8}\check@mathfonts
% needed to don't make the font size of the number of th equation change
%\makeatletter
%\def\tagform@#1{\maketag@@@{\normalsize(#1)\@@italiccorr}}
%\makeatother
$\left(\frac{2}{\eta}\right)^\text{th}$ moment is:
\begin{align}
\mathbb{E}^\text{Rayleigh}\left[g^\frac{2}{\eta}\right]=\frac{\left(\frac{2}{\eta}\right)!}{\mu^\frac{2}{\eta}}=\frac{\Gamma\left(\frac{2}{\eta}+1\right)}{\mu^\frac{2}{\eta}}
\end{align}
%\endgroup
The LT of the interference power in Rayleigh fading case becomes:
\begin{align}
L_I\left(s\right)=\exp{\left(-\pi\lambda\left(\frac{s}{\mu}\right)^\frac{2}{\eta}\frac{\pi\frac{2}{\eta}}{\sin{\left(\pi\frac{2}{\eta}\right)}}\right)}
\end{align}

\textbf{\textit{ LT of the interference in Rician fading}}: For $m=\frac{\left(K+1\right)^2}{2K+1}$ in the Nakagami fading case, we approximately have the Rician fading case with parameter $K$, where the $K$-factor is the ratio of the signal power in the dominant component (LOS-component) to the power of the other non-LOS components of an interference signal. For $m=\infty$, there is no fading.
\par In the case of a Rician channel, the fading power distribution can be written as $P_G(g)=\frac{1+K}{P_r}\exp\left(-K-\frac{1+K}{P_r}g\right)I_0\left(2\sqrt{\frac{K(1+K)g}{P_r}}\right)$, where $I_0(z)=\frac{1}{\pi}\int_{0}^{\pi}e^{z\cos(\theta)d\theta}$ is the modified Bessel function of first kind.
%It is worth reminding the reader that the exact PDF has never been addressed except for $\eta=4$\cite{haenggi2009interference}.
%The inner integral in \eqref{eq:PPP_laplace} can be solved for many values of $\eta$, but its ILT has never been obtained, except for $\eta=4$ .
%\vspace{-1em}
%\begingroup\makeatletter\def\f@size{8}\check@mathfonts
%\begin{table}[h]
%	\centering
%	\scriptsize
%	\begin{tabular}{ l|l|l|l }
%		\hline
%		Parameter & Definition & \begin{tabular}{@{}c@{}}Case 1 \\ (Fig. \ref{fig:figure1})\end{tabular} &\begin{tabular}{@{}c@{}}Case 2 \\ (Fig. \ref{fig:figure2})\end{tabular}\\ 
%		\hline
%		$\eta$& Path loss exponent& $3$ &$6$\\
%		$\mu$ & $1/\mu$ is the mean of the fading power &1 &1\\
%		$\lambda$ & Density of PPP & $2$ & not used\\
%		$\gamma$ & Scale parameter of L\'evy distribution & not used & $0.4$\\
%		$\beta$ & Index of scaled KWW & $2/3$ & $1/6$\\
%		\hline
%	\end{tabular}
%	\vspace{0.5em}
%	\caption{Simulation Parameters}
%	\label{tab:2}
%	\vspace{-1em}
%\end{table}
\subsection{Validation of the PDF Expressions}
\par First, for the case of $\eta=4$ and Rayleigh fading, it can be easily verified that the exact analytical expression of the interference PDF known in literature \cite[Equation~3.22]{haenggi2009interference} is obtained.

\par
\begin{figure}[h]
	\centering
	\includegraphics[width=1\linewidth]{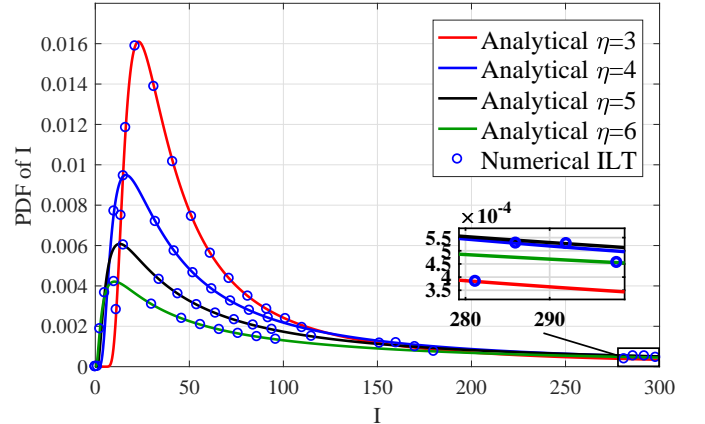}
	\caption{Interference power PDF for the Rayleigh fading case when $\mu=1$ and $\lambda=2$.}
	\label{fig:figure1}
	%\vspace{-1em}
\end{figure}

 To verify the different formulas in Table \ref{tab:1}, we use Talbot's method as a numerical solution for comparison. Talbot's method is one of the best approaches to compute the ILT by deforming the standard contour in the Bromwich inversion integral. It is widely used due to its accuracy, 
%\cite{abate2004multi}. %\cite{myTalbot}.
and the reader might refer to \cite[Section~3]{abate2004multi} for more details. In Fig. \ref{fig:figure1}, we provide and compare the results of the PDF of the interference power obtained analytically as in Table \ref{tab:1}, and numerically from the ILT Talbot's method for $\eta=3,4,5,$ and $6$. It is very clear that the analytical derivations results are exactly the same as those obtained by Talbots' method. %, $\lambda=2, \mu=1$. %, and different SINR thresholds $T$.
\textit{Hence, we claim that our approach tackles the general case of the aggregate interference PDF as long as the LT can be expressed as a KWW form, i.e., $exp(-ts^\beta)$.}
\vspace{-1em}
%Also, we can verify it by observing the pdf changes with respect to $T$, in which at very small $T$, for example at $T=-50$ $dB$, the interference becomes like a dirac function, while as $T$ tends to $\infty$ the stable distribution of interference is better observed. 
%The path loss exponent $\eta$ has a major impact, and acts as a stretching term which changes the interference that is experienced in the network. %We have evaluated the cases for $\eta=3,4,5,6$.
%\par The results have shown the accuracy of the derived expressions. %Here, one can also observe how the interference distribution changes with the threshold $T$. When $T$ increases, the probability to have small interference power is null, but it is flattened at higher interference values.
%One can observe how the interference power conditioned on $T$ changes, and how $T$ sets limits and bounds on the interference distribution. We can also observe the actual interference power pdf ($T=\infty$) and it is the one having the heaviest tail. %For instance, these results could be used for determining upper and lower bounds for the power allocations needed in a network.
%and the distribution mean is shifted toward higher values.
%Even though this is not the scope of the letter, but it is possible to use these results to analyze, improve and conclude on the network performance.
%The existence of power law which is used by the standard path loss model alongside with network model like PPP makes the interference power has a Stable distribution \cite{haenggi2009interference}.
\newline
\section{Application on the Probability of Coverage Derivations}
The probability of coverage is defined as the probability of the user receiving signal to interference and noise ratio (SINR) greater than a threshold value $T$. The threshold value determines the starting limit at which the received signal is considered useful so that the user can decode it. A basic SINR for a typical user in the previously defined model is given by:
\begin{align}
\text{SINR}=\frac{S}{I+\sigma^2}=\frac{hr^{-\eta}}{\sum _{i\in\mathrm{\Phi}}g_i{R_i}^{-\eta}+\sigma^2}
\end{align}
where $S,I,$ and $\sigma^2$ are the signal, interference, and noise powers respectively. $h$ is an arbitrary fading power channel coefficient experienced by the useful signal, and $r$ is the distance to the intended transmitter (e.g., serving base station), which is not necessarily the nearest transmitter, because in the analysis we are not assuming the existence of an interference exclusion region around the user.
\par When the PDF of the interference is known, a different approach from the one used in the literature \cite{6042301} can be used to obtain the probability of coverage. The importance here is mostly seen in obtaining the coverage when the useful signal experiences fading other than Rayleigh fading. Thus deriving the coverage cannot be done directly using the LT of the interference, because for fading scenarios other than Rayleigh, the LT of the interference does not come naturally in the coverage expression. The probability of coverage can be directly obtained as:
\begin{align}\label{eq:Pcov} 
p_c&=\mathbb{P}\left[SINR(r)>T\right]=\mathbb{P}\left[\frac{hr^{-\eta}}{\sigma^2+I_r}>T\right]\nonumber\\
&=\mathbb{P}\left[I_r<\frac{hr^{-\eta}}{T}-\sigma^2\right]=\mathbb{E}_h\left[\int_{0}^{\frac{hr^{-\eta}}{T}-\sigma^2}{f_I(x)dx}\right]\nonumber\\
&=\int_{0}^{\infty}{\left(\int_{0}^{\frac{hr^{-\eta}}{T}-\sigma^2}{f_I(x)dx}\right)f(h)dh}
\end{align}
where $T$ is the threshold to consider the received signal useful and $f_I(x)$ is the PDF of the interference derived previously. The PDF of $h$ depends on the type of fading experienced by the signal $S$. For example for $\eta=3$, when the interference signals experience Nakagami fading, the probability of coverage can be written as:
\begingroup\makeatletter\def\f@size{6}\check@mathfonts
% needed to don't make the font size of the number of th equation change
\makeatletter
\def\tagform@#1{\maketag@@@{\normalsize(#1)\@@italiccorr}}
\makeatother
\begin{align}\label{eq:coverageInterf}
p_c&=\int_{0}^{\infty}\Bigg(\int_{0}^{\frac{hr^{-\eta}}{T}-\sigma^2}\frac{\Gamma\left(\frac{2}{3}\right)t}{\sqrt{3}\pi I^{\frac{5}{3}}} \,_1F_1\left(\frac{5}{6};\frac{2}{3};-\frac{\frac{2^2}{3^3}t^3}{I^2}\right)+
\frac{\frac{2}{9}t^2}{\Gamma\left(\frac{2}{3}\right)I^\frac{7}{3}}
\,_1F_1\left(\frac{7}{6};\frac{4}{3};-\frac{\frac{2^2}{3^3}t^3}{I^2}\right)\nonumber\\
&dx\Bigg)f(h)dh=\int_{0}^{\infty}\xi\left(\frac{hr^{-\eta}}{T}-\sigma^2\right) f(h)dh
\end{align}
\endgroup
%\begin{strip}
with $t=\pi\lambda\frac{\mathrm{\Gamma}\left(m+\frac{2}{3}\right)}{\mathrm{\Gamma}\left(m\right)\left(\frac{m}{P_r}\right)^\frac{2}{3}}\mathrm{\Gamma}\left(1-\frac{2}{3}\right)s^\frac{2}{3}$
\begingroup\makeatletter\def\f@size{6}\check@mathfonts
\begin{align*}
&\begin{aligned}
\xi(U)=\frac{\Gamma \left(\frac{1}{3}\right)}{12 \pi  \Gamma \left(\frac{2}{3}\right)} \left(\xi_1(U)+\xi_2(U)+3 \sqrt{3} \Gamma \left(\frac{2}{3}\right)^2+\frac{6\ 2^{2/3} \pi ^{3/2}}{\Gamma \left(\frac{1}{6}\right)}\right)
\end{aligned}\\
&\begin{aligned}
\xi_1(U)=-\frac{6 \sqrt{3} \pi  \lambda \Gamma \left(\frac{2}{3}\right)^2 \Gamma \left(m+\frac{2}{3}\right) \left(\frac{P_r}{m U}\right)^{2/3} \, _2F_2\left(\frac{1}{3},\frac{5}{6};\frac{2}{3},\frac{4}{3};-\frac{4 \lambda^3 P_r^2 \pi ^3 \Gamma \left(\frac{1}{3}\right)^3 \Gamma \left(m+\frac{2}{3}\right)^3}{27 m^2 U^2 \Gamma (m)^3}\right)}{\Gamma (m)}
\end{aligned}\\
&\begin{aligned}
\xi_2(U)=-\frac{2 \pi ^3 \lambda^2 \Gamma \left(\frac{1}{3}\right) \Gamma \left(m+\frac{2}{3}\right)^2 \left(\frac{P_r}{m U}\right)^{4/3} \, _2F_2\left(\frac{2}{3},\frac{7}{6};\frac{4}{3},\frac{5}{3};-\frac{4 \lambda^3 P_r^2 \pi ^3 \Gamma \left(\frac{1}{3}\right)^3 \Gamma \left(m+\frac{2}{3}\right)^3}{27 m^2 U^2 \Gamma (m)^3}\right)}{\Gamma (m)^2}
\end{aligned}\\
\end{align*}
\endgroup
It is very clear that a similar approach can be derived for a different PLE $\eta$ value, or for different fading distributions on the interfering signals. The change in $\eta$ will be reflected in the value of $\beta$, and hence determining which equation to use from Table \ref{tab:1}. 
\par As a summary, the probability of coverage can be derived for different use cases as follows:
\begin{itemize}
\item Select the PLE $\eta$: this defines the value of $\beta$, hence indicating which formula to use from Table I.
\item Select the channel type of the interference from those provided in the previous section, e.g. Nakagami or Rayleigh. This determines the value of $m$, which in turn determines the value of $t$ used in \eqref{eq:coverageInterf}.
\item Plug-in the formula in the probability of coverage expressed in \eqref{eq:Pcov}.
\end{itemize}

\par
\begin{figure}[h]
	\centering
	\includegraphics[width=0.8\linewidth]{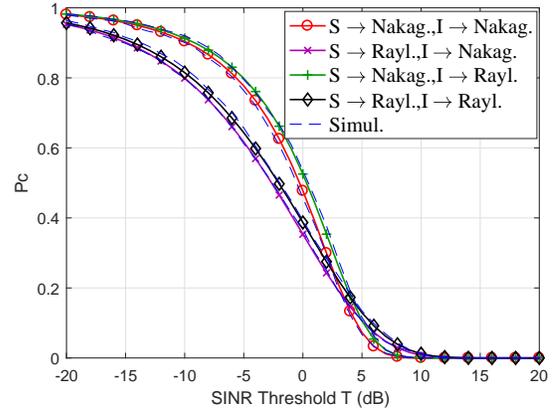}
	\caption{Probability of coverage when $\eta=3$}%, and when $S$ and $I$ experience either Rayleigh or Nakagami fading.}
	\label{fig:figure2}
	\vspace{-1em}
\end{figure}

\begin{figure}[h]
	\centering
	\includegraphics[width=0.8\linewidth]{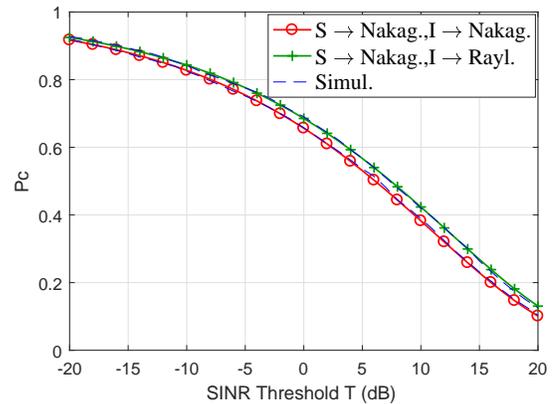}
	\caption{Probability of coverage when $\eta=6$}%, and when $S$ and $I$ experience either Rayleigh or Nakagami fading.}
	\label{fig:figure3}
	\vspace{-1em}
\end{figure}

 To verify our results, the probability of coverage obtained analytically has been compared to their respective Monte Carlo simulation results with 2000 trials. In Figs. \ref{fig:figure2} and \ref{fig:figure3}, an area of $40$x$40$ $km^2$ of a network whose transmitters are distributed as PPP has been considered. The typical user is placed at the origin and the serving transmitter (delivering useful signal) is placed at a specific distance $r=0.25$ $km$. The other system parameters are $\lambda=2,\mu=1,P_r=1,m=10,$ and $\sigma^2=0$. Four cases of channel fading have been considered: (1) both the useful signal $S$ and the interfering signals $I$ experience Nakagami fading, (2) both experience Rayleigh fading, (3) $S$ experiences Rayleigh fading while $I$ experiences Nakagami, and (4) the opposite of (3) is considered.
\par The results showed that the formulas are very useful in deriving the coverage for any type of fading. Similar results are obtained for other $\eta$ values as shown in Fig. \ref{fig:figure3}. For this network configuration, as expected, the highest achieved probability of coverage is when the signal experiences Nakagami fading and the interfering signals experience Rayleigh fading. This is because in the Rayleigh fading case there is no LOS between the receiver and the interfering source. The same can be verified for the other combinations obtained from the MonteCarlo simulations and the analytical formula using the PDF, where the lowest coverage was when the useful signal $S$ experiences Rayleigh fading and the interfering signals experience Nakagami. Moreover, for high PLE values (e.g. $\eta=6$), the decrease in the probability of coverage with respect to $T$ is smaller compared to that for low PLE (not strictly decreasing as seen for the $\eta=3$ case). It is true that both the useful signal and the interfering signals experience higher path loss at high PLE, but at the same time, the effect on the numerous interfering signals appears stronger.
%\par \hrulefill
%\end{strip}

\section{Conclusion}
We have shown that the distribution of the interference power in a stochastic geometry framework when the interfering sources are distributed according to PPP without an exclusion zone, does have exact closed forms, as long as its LT could be written as a KWW function. We have presented the closed form expressions for different path loss exponents for identical but arbitrary fading cases. Numerical ILT and Monte-Carlo simulations have shown the accuracy of the expressions, and the feasibility to calculate the coverage for any fading type.% This work will be extended in the future for other metrics, such as capacity.

% or


%\appendices

\appendix

\section{}
\label{Appendix}
The Laplace transform of interference power can be written as:
%\begingroup\makeatletter\def\f@size{6}\check@mathfonts
\begin{align*}
&L_I\left(s\right)=\mathbb{E}\left[e^{-sI}\right]=\mathbb{E}\left[\exp{\left(-s\sum_{i\in\mathrm{\Phi}}{g_i{R_i}^{-\eta}}\right)}\right]\\&=\mathbb{E}\left[\prod_{i\in\mathrm{\Phi}}\exp{\left(-sg_i{R_i}^{-\eta}\right)}\right]
\stackrel{\text{(a)}}{=}\mathbb{E}_\mathrm{\Phi}\left[\prod_{i\in\mathrm{\Phi}}{\mathbb{E}_{g_i}\exp{\left(-sg_i{R_i}^{-\eta}\right)}}\right]\\&\stackrel{\text{(b)}}{=}\exp{\left(-2\pi\lambda\int_{\mathbb{R}}\left(1-\mathbb{E}_g\exp{\left(-sgx^{-\eta}\right)}\right)xdx\right)}\\
&=\exp{\left(-2\pi\lambda\mathbb{E}_g\int_{\mathbb{R}}\left(1-\exp{\left(-sgx^{-\eta}\right)}\right)\text{xdx}\right)}\\
&\stackrel{\text{(c)}}{=}\exp{\left(-\frac{2\pi\lambda}{\eta}\mathbb{E}_g\int_{\mathbb{R}}{\left(1-\exp{\left(-\frac{sg}{y}\right)}\right)y^{\frac{2}{\eta}-1}\text{dy}}\right)}
\end{align*}
%\endgroup
where (a) follows from the independence of g, (b) from the Probability Generating Functional (PGFL) of PPP, and (c) from the change of variables $x^{-\eta}\rightarrow\frac{1}{y}$.\\
This expression resembles the definition of the $i^{th}$ moment of a random variable, which is  $\mathbb{E}\left[Y^i\right]=\int{iy^{i-1}\left(1-F(y)\right)dy}$, where $F(y)$ is the Cumulative Density Function (CDF) of $y$. Hence, it is the $\left(\frac{2}{\eta}\right)^{th}$ moment of the random variable with the CDF of $\exp{\left(-\frac{sg}{y}\right)}$. The CDF is for the random variable $Y^{-1}$ where $Y$ is exponential with mean $\frac{1}{sg}$. Consequently, 
%\begingroup\makeatletter\def\f@size{6}\check@mathfonts $\int_{\mathbb{R}}{\frac{2}{\eta}y^{\frac{2}{\eta}-1}\left(1-\exp{\left(-\frac{sg}{y}\right)}\right)dy}=\left(sg\right)^\frac{2}{\eta}\mathrm{\Gamma}\left(1-\frac{2}{\eta}\right)$. And the expression becomes:
\begin{align*}
L_I\left(s\right)=\exp{\left(-\pi\lambda\mathbb{E}\left[g^\frac{2}{\eta}\right]\mathrm{\Gamma}\left(1-\frac{2}{\eta}\right)s^\frac{2}{\eta}\right)}
\end{align*}
%\endgroup
%\\

\vspace{0.5em}
\section*{Acknowledgment}
The authors would like to acknowledge the generous support of the University Research Board (URB) of the American University of Beirut, Award $\#\ 103371$, Project $\#\ 24115$.
\vspace{0.5em}

\ifCLASSOPTIONcaptionsoff
  \newpage
\fi



%\bibliographystyle{IEEEtran}


%\newpage

\footnotesize
\bibliography{ICC_References}
\bibliographystyle{unsrt}


%\vfill


\end{document}